\documentclass[
	11pt,	% Set font size
	a4paper,	% Set paper size
	twocolumn
]{article}

\usepackage[a4paper,left=2.5cm,right=2.5cm,top=2.5cm,bottom=2.5cm]{geometry}

\usepackage[utf8]{inputenc}
\usepackage[english]{babel}
\usepackage[T1]{fontenc}
\usepackage{amsmath}
\usepackage{amsfonts}
\usepackage{amssymb}

\usepackage{graphicx}
\usepackage{float}
\usepackage{color}
\usepackage{hyperref}
\usepackage{subcaption}
\usepackage{trfsigns}
\usepackage{import}
\usepackage{multirow}
\usepackage{authblk}

\title{A Phase Synthesizer for Decorrelation  to Improve Acoustic Feedback Cancellation}
\author[1]{Klaus Linhard}
\author[2,*]{Philipp Bulling}

\affil[1]{Retired, formerly with Mercedes-Benz AG, Germany}
\affil[2]{Hochschule Esslingen, Germany}

\affil[*]{Corresponding author: \href{mailto:philipp.bulling@hs-esslingen.de}{philipp.bulling@hs-esslingen.de}}

\newcommand{\sub}[1]{_{\mathrm{#1}}} 						% Text tiefgestellt, nicht kursiv
\newcommand{\Exp}[1]{\mathrm{E}\!\left\{{#1}\right\}}			% Ewartungswert
\newcommand{\norm}[1]{\lVert {#1} \rVert}					% Norm
\newcommand{\vect}[1]{\boldsymbol{#1}}						% Vektoren fett
\newcommand{\matr}[1]{\textrm{\textbf{#1}}}					% Matrizen fett, roman

\newcommand{\h}{\vect{h}}									% Impulsantwort (Vektor)
\newcommand{\hEst}{\hat{\vect{h}}}							% Geschätzte Impulsantwort
\begin{document}
\maketitle

%========================================================================================
% Here, the main part starts
%========================================================================================
\section*{Abstract}
Undesired acoustic feedback is a known issue in communication systems, such as speech in-car communication, public address systems, or hearing aids. Without additional precautions, there is a high risk that the adaptive filter - intended to cancel the feedback path - also suppresses parts of the desired signal. One solution is to decorrelate the loudspeaker and microphone signals. In this work, we combine the two decorrelation approaches frequency shifting and phase modulation in a unified framework: a so-called \textit{phase synthesizer}, implemented in a discrete Fourier transform (DFT) filter bank. Furthermore, we extend the phase modulation technique using variable delay lines, as known from vibrato and chorus effects. We demonstrate the benefits of the proposed phase synthesizer using an example from speech in-car communication, employing an adaptive frequency-domain Kalman filter. Improvements in system stability, speech quality measured by perceptual evaluation of speech quality (PESQ) are presented.

\section{Introduction}
Adaptive acoustic feedback cancellation has many application fields, such as speech in-car communication~(e.g.,~\cite{Schmidt2006}), public address systems, and hearing aids. An extended overview of these applications and techniques is provided in~\cite{Watershoot2011}. In the following, we concentrate on microphone–loudspeaker systems with speech as the desired signal.

In the last 20 years, frequency-domain Kalman filters have been increasingly used to estimate the transfer function from the loudspeaker to the microphone~\cite{Enzner2006, Kuech2014}. Independent of whether a Kalman filter approach or another adaptive algorithm is used~(e.g.,~\cite{Linhard2021}), the adaptive filter $\hat{\h}$ aims to estimate the propagated loudspeaker signal as captured by the microphone signal $r$, in order to cancel the resulting acoustic feedback (echo).

\begin{figure}
  \centering
  %\import{images/}{MDF_Block.pdf_tex}
  \includegraphics{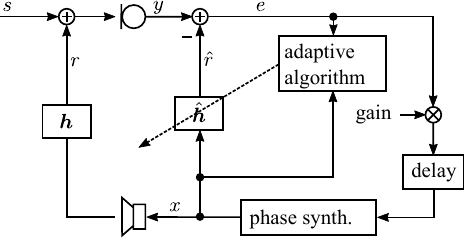}
  \caption{Acoustic feedback cancellation system with phase synthesizer included.}
  \label{fig:block_diagram}
\end{figure}

However, the microphone also serves as the input device for the original speech signal $s$. Since $s$ is correlated with the loudspeaker signal $x$, a fundamental problem arises: the adaptive filter may also cancel parts of the desired speech signal. A simplified block diagram of a feedback cancellation system is shown in Fig.~\ref{fig:block_diagram}.

\noindent
In Fig.~\ref{fig:block_diagram}, the proposed phase synthesizer is already integrated into the system. In addition to influencing the adaptive algorithm, the synthesizer also affects the loudspeaker signal and, consequently, the listener. Ideally, the applied phase modifications should decorrelate the signals $x$ and $s$, thereby facilitating convergence of the adaptive algorithm to the true room impulse response $\h$. At the same time, these modifications should not lead to a perceptible degradation in speech quality for the listener. 

Fig.~\ref{fig:block_diagram} also includes a gain parameter to control the loudspeaker level, as well as a processing delay. This delay arises due to the block-based processing required by the adaptive filter. In cases where long room impulse responses (e.g., 1000 samples or more) must be handled, efficient frequency-domain processing can be achieved using a multi-delay filter (MDF) structure with partitioned impulse response blocks~\cite{Soo1990}. The resulting system delay corresponds to the length of one partition. For example, if the total impulse response length is $N = 1024$ and we use $M = 4$ partitions, the partition delay is 256 samples.

It is important to emphasize that this processing-induced delay already contributes significantly to the decorrelation between $s$ and $x$, and is a key enabler of the proposed approach. Our phase synthesis method is also implemented in a block-wise manner, due to the overlap-add structure of the DFT filter bank. Segment length $N$ and overlap $L$ can be flexibly adjusted within certain bounds.

Our phase synthesizer realizes signal decorrelation through frequency shifting, phase modulation, variable time-delay lines, or combinations thereof. Decorrelation by means of frequency shifting has been demonstrated, for example, in~\cite{Withopf2014a,Bispo2012}, while phase modulation has been explored in~\cite{Herre2007,Guo2012}. In~\cite{Guo2012}, a combination of frequency shifting and phase modulation is implemented within a filter bank using complex-conjugated subbands.

Time-variable delay lines are well established in the domain of audio effects, such as chorus and vibrato~\cite{Zoelzer2011}, typically implemented in the time domain. Several interpolation techniques for realizing the required fractional delays are discussed in~\cite{Smith2022}. In our approach, the time-variable delay lines are implemented in the frequency domain and serve as a natural extension to frequency shifting and phase modulation. Fundamentally, all of these techniques represent different forms of phase modification, as only the phase of the signal is altered.

We implement these operations using a simple DFT filter bank rather than a more complex subband structure as in~\cite{Guo2012}, thereby increasing the method's practicality and applicability. Within our DFT-based framework, frequency information cannot be transferred between frequency bins; we restrict modifications to phase changes within the same bin. In contrast to approaches like~\cite{Withopf2014a}, we do not perform frequency-bin shifts.

For a sampling rate of $f_a = 16\textrm{\,kHz}$ and a DFT length of $N = 256$, the frequency resolution is $f_a/N = 62.5\textrm{\,Hz}$. In practice, frequency shifts should remain well below this upper limit. The introduced error depends on the chosen segment overlap and window function. As we will show, this constraint is acceptable for the targeted application.

Other approaches for signal decorrelation have also been proposed in the literature. These include whitening the signal solely for the adaptation process using linear prediction, as in~\cite{Bernardi2017}, applying non-linear signal distortions~\cite{Morgan2001, Bispo2012}, or injecting artificial noise~\cite{Valin2006}. Such techniques may be employed in addition to the proposed phase synthesizer. However, these methods are beyond the scope of this paper and will not be discussed further.

In the following sections, we first introduce an objective speech quality measure, the Perceptual Evaluation of Speech Quality (PESQ). We then analyze the bias problem resulting from the previously discussed correlation between the loudspeaker signal $x$ and the desired speech signal $s$. A subsequent section provides a detailed description of the phase synthesizer and presents PESQ-based speech quality results. 

Thereafter, we evaluate the performance of the proposed approach in the context of an adaptive Kalman filter \cite{Kuech2014}, focusing on PESQ scores, convergence speed, and final misadjustment. Finally, we summarize the findings and draw conclusions in the concluding section.

\section{Speech Quality with PESQ}
Subjective listening tests are inherently time-consuming, as they require participation from multiple listeners and the evaluation of a large amount of audio data. During the development phase, it is therefore more practical to rely on so-called objective speech quality measures.

We considered two widely used metrics: \textit{Perceptual Evaluation of Speech Quality} (PESQ)~\cite{Hu2006} and the \textit{Virtual Speech Quality Objective Listener} (ViSQOL)~\cite{Hines2012}. Both methods are available as MATLAB implementations. PESQ and ViSQOL estimate the Mean Opinion Score (MOS), which reflects perceived speech quality on a scale from 1 to 5: 1 (bad/very annoying), 2 (poor/annoying), 3 (fair/slightly annoying), 4 (good/perceptible but not annoying), and 5 (excellent/imperceptible).

After comparing both methods, we decided to report only PESQ results in this paper. In our experiments, ViSQOL yielded relatively small MOS differences for the types of distortions under consideration, whereas PESQ was more sensitive to these variations. A more detailed comparison between PESQ and ViSQOL is provided in~\cite{Hines2013}.

\section{Bias Problem and Decorrelation}

Deriving the least means squared error 
\begin{equation}
\Exp{e^2}\rightarrow\min,
\end{equation}
where $\Exp{\cdot}$ denotes the expected value, with respect to the estimate  $\hEst$ finally results in the optimum impulse response estimate
\begin{equation}
\hEst\sub{opt} = \h + \h\sub{bias} = \h + \matr{R}^{-1}_{xx} \vect{r}_{xs},
\label{eq:ir_opt}
\end{equation}
as shown in~\cite{Puder2004}. The vector $\hEst\sub{opt}$ is composed of two parts: the true impulse response vector $\h$ of the room and the bias impulse response $\h\sub{bias}$. The matrix $\matr{R}_{xx}$ denotes the auto-correlation matrix of the vector $\vect{x}$, while the cross-correlation between the vectors $\vect{x}$ and $\vect{s}$ is represented by the vector $\vect{r}_{xs}$. 

Assuming the impulse responses have length $N$, the vectors have dimensions $(N \times 1)$ and the matrix has dimensions $(N \times N)$. The second component, $\h\sub{bias}$, acts as a predictor, i.e., it represents the predictable portion of $\vect{s}$ using $\vect{x}$ as input. A high cross-correlation vector $\vect{r}_{xs}$ (in the non-causal part) indicates strong predictability and is therefore associated with reduced performance in our application.

We evaluate how a fixed delay $D$ affects the cross-correlation and reduces the resulting bias. The prediction estimate is obtained by convolving the delayed signal $s(k-D)$ with $\h\sub{bias}$, where $k$ denotes discrete time. This yields the prediction error
\begin{equation}
e(k) = s(k) - \h\sub{bias} * x(k),
\end{equation}
where the input signal $x(k)$ is the delayed version of the speech signal, i.e., $x(k) = s(k-D)$. 

We define the prediction gain $g_p$ as the ratio of the variances of $s$ and the prediction error $e$:
\begin{equation}
g_p = \frac{\sigma^2_s}{\sigma^2_e}.
\label{eq:PredGain}
\end{equation}

A phonetically balanced speech sentence may be used to calculate the prediction gain $g_p$. It is known that prediction of the noisy speech components, primarily consonants, is poor or even impossible. In contrast, prediction is mainly effective in the voiced parts, namely the vowels.

To isolate this effect, we created a second speech example consisting solely of vowels: the five German vowels \textit{a–e–i–o–u}, each about 1\,sec long and combined into a vowel sequence of 5\,sec duration. Basically, we could compute $\h\sub{bias}$ via the inverse matrix solution in Eq.~(\ref{eq:ir_opt}), and use the causal part of $h\sub{bias}$.

Since our later application uses a Kalman-based version of a frequency domain least mean squares algorithm (FLMS, \cite{Soo1990,Kuech2014}), we chose to solve Eq.~(\ref{eq:ir_opt}) using a standard FLMS with one partition ($M=1$) and a normalized step size of $\alpha=0.4$.

\begin{figure}
\centering
\begin{subfigure}{\linewidth}
  \includegraphics[width=\linewidth]{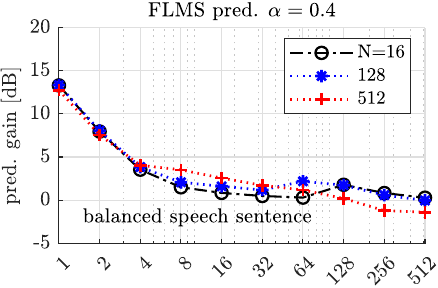}
\end{subfigure}
\begin{subfigure}{\linewidth}
  \includegraphics[width=\linewidth]{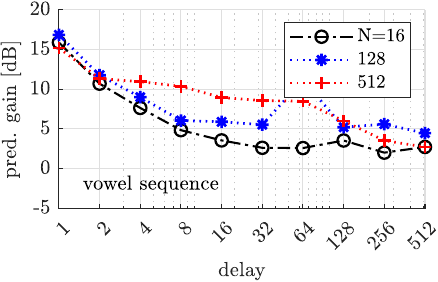}
\end{subfigure} 
  \caption{Prediction gain $g_p$ vs. delay $D$, FLMS solution with $N= 16,128,512$.}
  \label{fig:PredGain_FLMS}
\end{figure}

We calculated the prediction gain $g_p$ for different predictor lengths $N$ and various delay values $D$. Fig.~\ref{fig:PredGain_FLMS} presents the resulting prediction gains. The upper plot shows results for a phonetically balanced sentence spoken by a male speaker, while the plot on the bottom illustrates results from the vowel sequence \textit{a–e–i–o–u} (male speaker).
\noindent 
% BiasMixSpeechDelay_FLMS_pack.m
% BiasVoicedDelay_FLMS_pack.m
For the phonetically balanced case, the prediction gain drops to nearly 0 already after a short delay of only a few samples. In contrast, for the vowel-only signal, we observe that, for example, at a delay of $D=64$, the prediction gain remains around 10\,dB. A gain of 10\,dB implies that the prediction error accounts for approximately 30\,\%, meaning about 70\,\% of the signal can still be predicted.

In our later application, we will require block processing with, for example, $N=512$ and a delay of 256 samples. From this experiment, we conclude that a fixed delay introduces a significant decorrelation effect already with short delays. However, in the voiced speech parts, a residual correlation may still exist, which can degrade the overall performance.

\section{Phase Synthesizer}
In the frequency domain, the correspondence to the time segment $x_l(k)$ is given by
\begin{equation}
\vect{X}(l,n) = |\vect{X}(l,n)| \cdot e^{\mathrm{j} \cdot \varphi(n,l)},
\label{eq:PhaseExpon}
\end{equation}
where $n$ denotes the discrete frequency bin and $l$ the discrete frame index.

Using a DFT filter bank results in block processing with frame index $l$, which corresponds to a time interval of $L$ samples. Typically, the frame shift $L$ equals $N/2$ or $N/4$, where $N$ is the segment length and also the DFT size. Before applying the DFT and after the inverse DFT (IDFT), the segments are multiplied by a normalized Hanning window $\vect{w}$ of length $N$. For $L = N/2$, the window is defined as
\begin{equation}
\vect{w} = \sqrt{\frac{2L}{N}} \cdot \sqrt{\mathrm{hann}(N)},
\end{equation}
and for $L = N/4$ as
\begin{equation}
\vect{w} = 2 \cdot \sqrt{\frac{L}{1.5N}} \cdot \mathrm{hann}(N).
\end{equation}
The frame index $l$ is connected to time
\begin{equation}
t = \frac{L\cdot l}{f_a}.
\end{equation}
A frequency shift of $f_s$ can be realized by adding the phase increment
\begin{equation}
\varphi_{\mathrm{add}} = 2\pi \frac{f_s}{f_a} \cdot L \cdot l
\label{eq:PhaseAddShift}
\end{equation}
to the phase component $\varphi(n,l)$ of Eq.\,\ref{eq:PhaseExpon}. A sampling frequency of $f_a = 16\textrm{\,kHz}$ is used throughout the paper.

To implement phase modulation, we express Eq.~(\ref{eq:PhaseAddShift}) in a periodic form as
\begin{equation}
\varphi_{\mathrm{add}} = a \cdot \sin\left( 2\pi \frac{f_p}{f_a} \cdot L \cdot l \right),
\end{equation}
where $f_p$ denotes the modulation frequency and $a$ the modulation amplitude. The sine function may be replaced by any other periodic function or even by low-pass filtered random noise.

Since the phase increases with each phase addition, it is advisable to apply a modulo operation to confine the phase values within the interval $[-\pi, \pi]$.

We now extend the phase modulation approach to realize time-varying delay lines. Variable delay lines are commonly employed to produce well-known audio effects such as vibrato and chorus~\cite{Zoelzer2011}. Physically, vibrato corresponds to the periodic modulation of the pitch in a singing voice or a musical instrument (e.g., violin). Chorus, on the other hand, arises from the non-synchronous onset and slight pitch variations of multiple singers or instruments, exhibiting a more stochastic character. Typically, the chorus effect is created by combining the outputs of several delay lines, each modulated by a different low-pass filtered noise signal.

We focus on the vibrato effect by introducing a periodically modulated phase with a linear slope over the frequency range, expressed as
\begin{equation}
\varphi_{\mathrm{add}} = \frac{2n}{N} \cdot a \cdot \sin\left( 2\pi \frac{f_p}{f_a} \cdot L \cdot l \right),
\label{eq:PhaseAddVibrato}
\end{equation}
where $n = 0, 1, 2, \ldots, N/2$. At the lowest frequency bin $n=0$, the phase addition is zero, while at $n = N/2$, it attains its maximum magnitude
\begin{equation}
\mathrm{max}\left\{|\varphi_{\mathrm{add}}(n)|_{n=N/2}\right\} = a.
\end{equation}

By applying the time-shift correspondence $n = N/2$
\begin{equation}
x(k - k_s) \;\laplace\; e^{-\mathrm{j} k_s n 2\pi / N} \cdot X(n),
\end{equation}
the phase at frequency $n$ corresponds to $e^{-\mathrm{j} k_s \pi}$. Thus, a time shift of $k_s$ samples induces a maximum phase addition of $k_s \pi$.

Substituting $a = -k_s \pi$ into Eq.~\eqref{eq:PhaseAddVibrato} yields
\begin{equation}
\varphi_{\mathrm{add}} = -\frac{2n}{N} \cdot k_s \pi \cdot \sin\left( 2\pi \frac{f_p}{f_a} \cdot L \cdot l \right).
\end{equation}

\begin{figure}
\centering
\begin{subfigure}{\linewidth}
  \includegraphics[width=\linewidth]{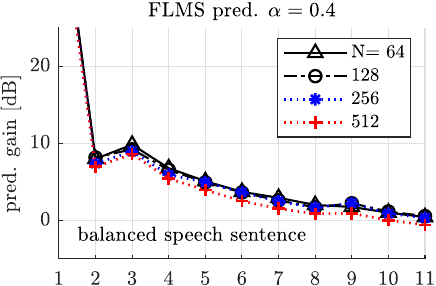}
\end{subfigure}
\begin{subfigure}{\linewidth}
  \includegraphics[width=\linewidth]{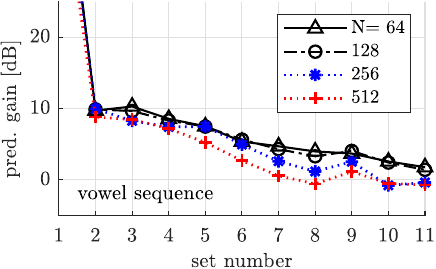}
\end{subfigure} 
  \caption{Prediction gain of phase synthesizer with different parameter sets (1-11) and $N= 64,128,256,512$.}
  \label{fig:PhaseSyn}
\end{figure}\noindent
% BiasMixSpeechPSy_FLMS_pack030.m
% BiasVoicedPSy_PESQ_pack030.m
Fig.~\ref{fig:PhaseSyn} presents the results of the phase synthesizer evaluated using eleven distinct parameter sets (labeled 1 to 11), and summarized in Table~\ref{tab:ParamSets}. For a better comparison, we keep the subband numbering of \cite{Guo2012}, with subbands of bandwidth 312.5\,Hz, each. E.g., for subband number 3 we assume a center frequency of $3\cdot 312.5\mathrm{\,Hz}=937.5\mathrm{\,Hz}$. For our DFT filter bank implementation, we use the center frequencies and perform linear interpolation to $f_a/2$ to obtain these phase modifications for all intermediate frequency bins.

\begin{table}[t]
\centering
\caption{Parameter sets of the phase synthesizer.}
\begin{tabular}{|c|c|c|c|c|c|l|}
\hline
\multirow{2}*{Set} & \multicolumn{5}{c|}{Subbands} & \multirow{2}*{Description} \\ 
\cline{2-6}
 & 0-3 & 4 & 5 & 6 & $\geq 7$ &  \\ 
\hline 
\hline 
1) & \multicolumn{5}{c|}{no modification} &  \\ 
\hline 
2) & 0 & 10 & 10 & 10 & 10 & [Hz]\\ 
\hline 
3) & .11 & .22 & .39 & .5 & 1 & $[\pi\textrm{rad}]$; 10\,Hz\\ 
\hline 
4) & \multicolumn{5}{c|}{combine sets 2) and 3)} &  \\ 
\hline 
5) & 0 & \multicolumn{3}{c|}{interp. to $f_a/2$} & 8 & $[\pi\textrm{rad}]$; 1\,Hz\\ 
\hline 
6) & 0 & \multicolumn{3}{c|}{''} & 16 & $[\pi\textrm{rad}]$; 1\,Hz\\ 
\hline 
7) & 0 & \multicolumn{3}{c|}{''} & 16 & $[\pi\textrm{rad}]$; 2\,Hz\\ 
\hline 
8) & 0 & \multicolumn{3}{c|}{''} & 16 & $[\pi\textrm{rad}]$; 3\,Hz\\ 
\hline 
9) & 0 & \multicolumn{3}{c|}{''} & 32 & $[\pi\textrm{rad}]$; 1\,Hz\\ 
\hline 
10) & 0 & \multicolumn{3}{c|}{''} & 32 & $[\pi\textrm{rad}]$; 2\,Hz\\ 
\hline 
11) & 0 & \multicolumn{3}{c|}{''} & 32 & $[\pi\textrm{rad}]$; 3\,Hz\\ 
\hline 
\end{tabular} 
\label{tab:ParamSets}
\end{table}
Parameter sets 2), 3), and 4) correspond to those reported in~\cite{Guo2012}, whereas sets 5) to 11) represent examples of the variable delay lines introduced in this work. Set 2) uses a frequency shift of 10\,Hz. Set 3) is a setting for phase modulation, using a sine wave with modulation frequency 10\,Hz. Set 4) is the combination of sets 2) and 3). As noted in~\cite{Bispo2012},~\cite{Herre2007}, and~\cite{Guo2012}, frequency shifting and phase modulation should generally be avoided in the lower frequency range (below 2\,kHz) to preserve speech quality. However, small frequency shifts in the higher frequency range are typically imperceptible to listeners. 

For the proposed variable delay line (commonly referred to as vibrato in audio effect applications), the phase modulation amplitude begins at zero and increases linearly up to values of $8\pi$, $16\pi$, or $32\pi$ at the highest frequency $f_a/2$ in our parameter sets. For instance, a modulation amplitude of $16\pi$ corresponds to a maximum variable delay of approximately $\pm 1$\,ms. In practice, the modulation amplitude may be limited, e.g., to a value of $\pm\pi$. However, in this work we did not apply such a limit in order to maintain the analogy to the variable delay line. The delay modulation was driven by a sinusoidal signal at either 1\,Hz, 2\,Hz, or 3\,Hz.

Fig.~\ref{fig:PhaseSyn} shows the prediction gains for the phonetically balanced sentence (top), and the corresponding results for the vowel sequence (bottom), both plotted against the parameter set numbers defined in Table~\ref{tab:ParamSets}. Set 1) corresponds to the case without any modification. Here, prediction is perfect, resulting in a very high gain that exceeds the graphical scale of the figure. For sets 2), 3), and 4), the prediction gain is approximately 8 to 10\,dB. For the delay lines corresponding to sets 5) through 11), the prediction gain ranges between 6\,dB and 0\,dB. 

In the case of the vowel sequence, the prediction gain is slightly higher compared to the balanced sentence, reflecting the more predictable structure of voiced segments. The results shown exclude the inherent delay introduced by the block processing of our DFT filter bank, which was compensated prior to performing the prediction. If this processing delay were included, the prediction gains would be close to 0\,dB for all parameter sets.

\begin{figure}
\centering
\includegraphics[width=\linewidth]{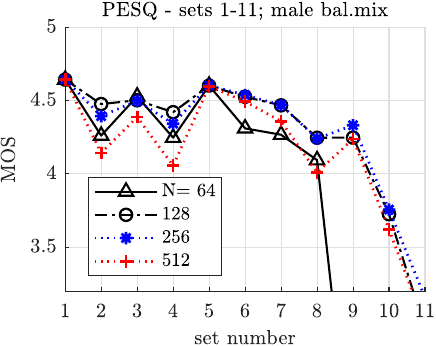}
  \caption{Speech quality MOS with PESQ for phase synthesizer with different parameter sets 1)-11) and $N= 64,128,256,512$.}
  \label{fig:PhaseSynPESQ}
\end{figure}\noindent 
%BiasMixSpeechPSy_PESQ_pack030.m
Fig.~\ref{fig:PhaseSynPESQ} shows the PESQ results for the phonetically balanced sentence. For parameter sets numbered 9) and higher, which apply more extensive modifications, the MOS value drops below 4. An interesting example is set 4), from~\cite{Guo2012}. When comparing set 4) with the variable delay lines of sets 6) and 7), we observe that although sets 6) and 7) exhibit lower prediction gains, they achieve higher PESQ values. Set 9) also indicates promising performance (if we exclude $N=64$). These combinations—low prediction gain with high PESQ—are particularly desirable and are selected for further evaluation in the feedback experiments.

\section{Kalman Feedback Cancellation with the Phase Synthesizer}
In Fig.~\ref{fig:block_diagram}, the phase synthesizer is already integrated into the structure of an acoustic feedback cancellation (AFC) system. We now present the improvements we achieved.

The room impulse response used in our evaluation corresponds to a typical in-car speech communication scenario, with a length of 1024 samples at a sampling frequency of $f_a=16$\,kHz, for the first test. The speech signal consists of a phonetically balanced sentence spoken by a male speaker and has a total duration of 42\,s. The acoustic coupling between the loudspeaker and microphone was adjusted such that the level of the room signal $r$ at the microphone position was approximately 10\,dB below the input speech signal $s$ (i.e., coupling gain $\approx -10$\,dB, at loop gain 0\,dB).

The Kalman filter was realized using a multi-delay filter (MDF) structure with $M=4$ partitions, each of length $N=512$. The Kalman filter parameter $A$ was set to $A = 0.99999$~\cite{Kuech2014}.

Due to the block-based processing of the MDF Kalman structure, an inherent delay of 256 samples is introduced. The phase synthesizer was implemented as an add-on module without further optimization, as depicted in Fig.~\ref{fig:block_diagram}. It uses a DFT filter bank with $N=256$ and half-overlapping blocks. In our filter bank, this frame shift $L=128$ results in an additional delay of 128 samples.

We present results for parameter sets 1), 4), 6), and 9) (see Table\,\ref{tab:ParamSets}). Set 1) corresponds to the baseline with phase modification disabled (introducing only the processing delay), set 4) employs parameters from~\cite{Guo2012}, and sets 6) and 9) use the proposed variable delay lines modulated with a 1\,Hz sinusoidal signal. Set 6) corresponds to a maximum delay of $\pm 1$\,ms, and set 9) to $\pm 2$\,ms. 

\begin{figure}
  \includegraphics[width=\linewidth]{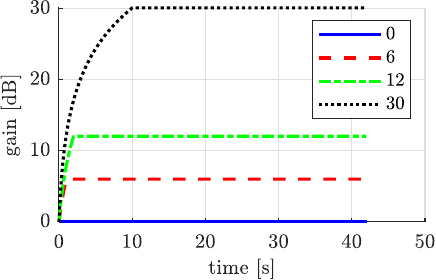}
  \centering
  \caption{Gain ramp at adaptation start and final gain.}
  \label{fig:GainRamp}
\end{figure}\noindent
%plot_g_sav.m}
The loop gain was gradually increased at the beginning of the experiment to simulate typical AFC conditions. It was set to 0, 6, 12, and 30\,dB, as illustrated in Fig.~\ref{fig:GainRamp}.

\begin{figure}
\centering
\begin{subfigure}{\linewidth}
  \includegraphics[width=\linewidth]{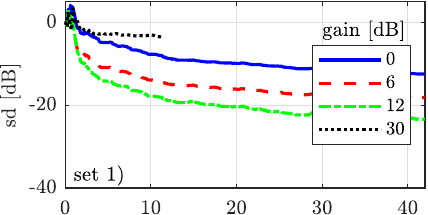}
\end{subfigure}
\begin{subfigure}{\linewidth}
  \includegraphics[width=\linewidth]{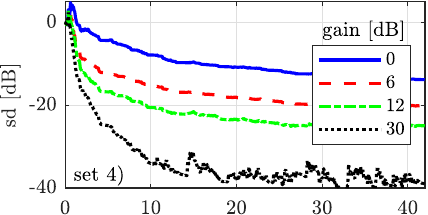}
\end{subfigure}
\begin{subfigure}{\linewidth}
  \includegraphics[width=\linewidth]{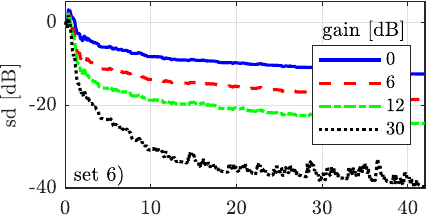}
\end{subfigure}
\begin{subfigure}{\linewidth}
  \includegraphics[width=\linewidth]{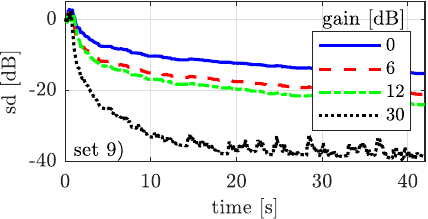}
\end{subfigure}
  \caption{Convergence of system distance for different parameter settings and loop gains. Top: Set 1) no phase modification. Mid-top:  Set 4) of~\cite{Guo2012}. Mid-bottom: Set 6) vibrato 1\,Hz sine, $\pm 1$\,msec. Bottom: Set 9) vibrato 1\,Hz sine, $\pm 2$\,msec.}
  \label{fig:ConvergenceResults}
\end{figure} 
%FBC_MDF_KalmanPredStdCopRatPSy02_pack02_x.m
Fig.~\ref{fig:ConvergenceResults} shows the system distance $\textrm{sd}(l)$ for the parameter sets 1), 4), 6), and 9). The system distance is computed for each time block index $l$, based on the true room impulse response vector $\h_0$ and the estimated impulse response $\hEst_l$
\begin{equation}
\textrm{sd}(l) = \norm{\h_0 - \hEst_l} / \norm{\h_0},
\end{equation}
where $\norm{\cdot}$ denotes the L2-norm.

The upper plot in Fig.~\ref{fig:ConvergenceResults} shows the system distance curves without any phase modifications, but including the delay caused by block processing. Note that in the 30\,dB gain case, the system becomes unstable after approximately 10\,s, causing the curve to terminate as $\textrm{sd} \rightarrow \infty$.

The second plot shows the performance of the combined frequency shift and phase modulation method proposed by~\cite{Guo2012}. The third plot presents the results for the variable delay line implementation with a sinusoidal modulation of 1\,Hz and a maximum delay of $\pm 1$\,ms. Finally, the bottom plot shows the same structure but with an increased delay of $\pm 2$\,ms at 1\,Hz.

\begin{figure}
  \includegraphics[width=\linewidth]{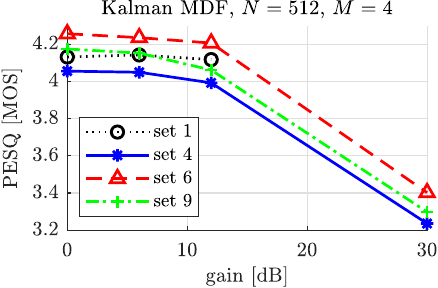}
  \centering
  \caption{MOS vs. final gain and parameter setting: 1) no phase modification; 4) acc.~\cite{Guo2012}; 6) vibrato $\pm 1$\,msec; 9) vibrato $\pm 2$\,msec.}
  \label{fig:PESQResults}
\end{figure}\noindent
%plot_PESQ_PSy02.m
The MOS values for the four selected parameter sets are presented in Fig.~\ref{fig:PESQResults}. These values are derived using PESQ after 20\,s of processing time, by which point the system has reached convergence. Parameter set 1) represents the baseline without any phase modification. At a feedback gain of 30\,dB, set 1) became unstable; hence, no PESQ value is reported for this case.

The MOS performance of sets 4), 6), and 9) is similar, with a slight advantage observed for the proposed variable delay line configurations (sets 6 and 9). 

From the first system tests with one speech file, we may summarize that the system distance of the parameter sets 4), 6) and 9) perform similar, but MOS is higher for 6) and 9).

\section{Results with speech and impulse response databases}
In the previous sections, we consistently used the same phonetically balanced sentence (male voice) and, in some cases, a vowel sequence to evaluate the system. However, to obtain more robust and generalizable results, it is now necessary to include a significantly larger and more diverse set of test data. We used two publicly available databases: the Lombard speech database in German~\cite{Soloducha2016} and the Automotive Noise and Impulse Response (ANIR) corpus~\cite{Huebschen2022}. From the speech database, we selected recordings from two female and two male speakers, each providing two sentences. Only the Lombard-free speech was used, as the focus of this work is not on the Lombard effect. Since each sentence has a duration of approximately 6 to 10\,sec, we repeated each sentence to generate longer sequences of 42\,sec.

From the ANIR corpus, we selected three different impulse responses. Specifically, we used the impulse response from the headliner driver microphone (entry 1 in the corpus) to the door speaker of the driver, and to the left and right side door loudspeakers in the rear of the car (entries 18, 20, and 21 in the corpus). The combination of 8 speech signals and 3 impulse responses yields a total of 24 test samples. Considering the 4 feedback gain settings (0, 6, 12, and 30\,dB), we obtained a total of 96 speech samples for evaluation (to be multiplied with the number of parameter sets 4), 6) and 9)).

The acoustic coupling between loudspeaker and microphone was again set to -10\,dB. To ensure a more natural frequency balance in playback, we applied a simple low-frequency equalization to the ANIR in-car impulse responses (recorded in a Mercedes van), as the original responses exhibited an excessive bass component.

To evaluate the performance, we present three types of results: MOS (Mean Opinion Score), early system distance, and late system distance. The early system distance provides insight into the convergence speed of the adaptive algorithm, while the late system distance reflects its steady-state accuracy. The early distance is computed as the average over the interval $[4,6]$\,sec, and the late distance as the average over the interval $[20,41]$\,sec. The MOS value is calculated based on the last complete sentence within the $[20,41]$\,sec interval.

For meaningful averaging, we grouped the results into clusters. We observed that the three different impulse responses produced very similar outcomes, allowing us to average them together. Furthermore, speech samples from male speakers showed similar performance, forming a consistent male cluster. The same held true for female speakers, who were grouped into a separate female cluster. Results are also shown separately for the different loop gain settings.

\begin{figure}
  \includegraphics[width=\linewidth]{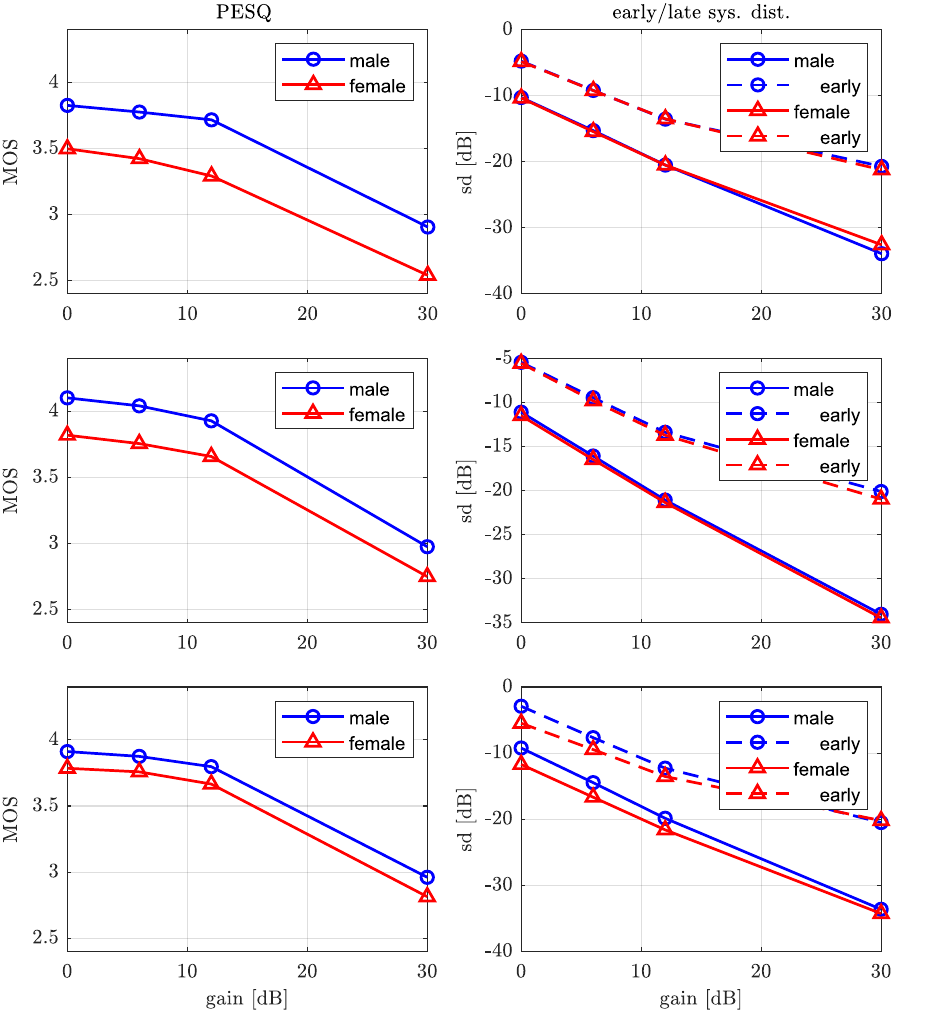}
  \centering
  \caption{Performance summary for the processing of about 100 different speech samples with 3 different parameter sets. Top: Set 4) phase modulation and frequency shift according to~\cite{Guo2012}. Mid: Set 6) variable delay line $\pm 1$\,msec, 1\,Hz. Bottom: Set 9) variable delay line $\pm 2$\,msec, 1\,Hz.}
  \label{fig:PerformanceSummary}
\end{figure}

Fig.~\ref{fig:PerformanceSummary} presents a performance summary for three different parameter sets. The optimal configuration is characterized by the highest MOS combined with the lowest early and late system distances. While system distances showed no significant differences between parameter sets, the MOS values indicate a clear trend: the variable delay lines with a maximum delay of $\pm 1$\,msec yielded the best overall speech quality in this summary (1\,Hz modulation).

\noindent

\section{Conclusion}
We proposed a phase synthesizer as a flexible and efficient tool to achieve decorrelation between the loudspeaker and microphone signals in acoustic feedback cancellation systems. The synthesizer is implemented as a DFT filter bank with overlapping, windowed segments and can be seamlessly integrated as an add-on module to existing frequency-domain adaptive algorithms, such as the Kalman filter-based feedback canceller.

While phase modulation and frequency shifting in the higher frequency range are established techniques for inducing decorrelation, we extended these methods by introducing a time-varying delay line, an effect analogous to vibrato or chorus in audio processing. This natural and perceptually motivated modulation strategy enhances decorrelation while preserving speech quality.

Our evaluation, based on the objective speech quality metric PESQ and publicly available databases, confirms the effectiveness of the approach. In addition to PESQ, we employed early and late system distance metrics to assess convergence behavior and steady-state accuracy. The results demonstrate that the phase synthesizer, particularly with the variable delay line, provides a robust and perceptually transparent decorrelation mechanism that improves upon existing solutions.

\section*{Data Availability Statement}
The audio and impulse response data used in this work come from publicly available resources. The Lombard speech recordings \cite{Soloducha2016} are available on Zenodo (\url{https://zenodo.org/records/48713}). The ANIR in-car impulse response corpus \cite{Huebschen2022} is available from the Digital Signal Processing and System Theory Group at Kiel University (\url{https://dss-kiel.de/index.php/media-center/data-bases/anir-corpus}). All datasets are accessible to the public under the terms specified by their respective providers. No proprietary or restricted data were used. % Chapter in separate file!

\bibliography{iccBib} 
\bibliographystyle{ieeetr}

\end{document}